# The contribution of star scientists to overall sex differences in research productivity[1]

*Giovanni Abramo[a,b*], Ciriaco Andrea D'Angelo[a] and Alessandro Caprasecca[a]*

[a] Laboratory for Studies of Research and Technology Transfer
Department of Management, School of Engineering
University of Rome "Tor Vergata"

[b] Italian Research Council

**Abstract**

The state of the art on the issue of sex differences in research efficiency agrees in recognizing higher performances for males, however there are divergences in explaining the possible causes. One of the causes advanced is that there are sex differences in the availability of aptitude at the "high end". By comparing sex differences in concentration and performance of Italian academic star scientists to the case in the population complement, this work aims to verify if star, or "high-end", scientists play a preponderant role in determining higher performance among males. The study reveals the existence of a greater relative concentration of males among star scientists, as well as a performance gap between male and female star scientists that is greater than for the rest of the population. In the latter subpopulation the performance gap between the two sexs is seen as truly marginal.

---

* **Corresponding author**, Università degli Studi di Roma "Tor Vergata", Dipartimento di Ingegneria dell'Impresa, Via del Politecnico 1, 00133 Roma, ITALY; Tel. +39 06 72597362, Fax +39 06 72597305, abramo@disp.uniroma2.it

# 1. Introduction

The subject of performance differences for men and women engaged in scientific research is one of great interest, having attracted the attention of scientometrists, sociologists of science and cognitivists for more than the past twenty years. Literature on the theme seems to agree that male researchers do indeed publish more than women (Mauleón and Bordons, 2006; Lee and Bozeman, 2005; Xie and Shauman, 2004 and 1998; Long, 1992 and 1987; Cole and Zuckerman, 1984; Fox, 1983).

Various studies have attempted to probe the reasons for this differential (Leahey, 2006; Xie and Shauman, 2003). In particular, psycho-cognitive studies exploring verbal (Hyde and Linn, 1988), spatial (Linn and Peterson, 1985; Voyer et al., 1995) and mathematical abilities (Hyde et al., 1990) reveal that, for these abilities, sex differences are found only in few dimensions and, where they do occur, are very limited. This has raised the so-called "productivity puzzle", which several science sociologists have attempted to resolve (Fox and Mohapatra, 2007; Palomba & Menniti, 2001; Etzkowitz et al., 2000). When identifying the fundamental drivers of performance, these sociologists have noted the importance of marriage status and stage of the family life cycle (see especially Fox, 2005 and Stack, 2004). The greater difficulty that women experience in balancing professional and family life is certainly determinant of performance (Stack, 2004). However some studies on the subject also emphasize that the presence of children, particularly of pre-school age, while implying a reduction of the total time that can be dedicated to work (a reduction that is typically more significant for women), also constitutes a psychological incentive and motivational driver towards higher efficiency in the allocation and use of time available (Prpic, 2002; Fox, 2005). The question of sex difference in research productivity has always been highly sensitive, as witnessed by the extreme example, in 2006, of Lawrence Summers being dismissed from his position as president of Harvard University. Although other motives were also expressed, the dismissal came following Summer's public statements on the possible causes of disparities in the presence of women in high-end scientific professions (Summers, 2005). Summers suggested three broad hypotheses about the sources of the very substantial disparities. The cause which Summers indicated as being least important concerned inequality in opportunity for the two sexes, which he termed "different socialization and patterns of discrimination in a search". Another potential cause mentioned was the "high-powered job hypothesis", being that mothers have a hard time reaching the top in jobs where people work long hours and put everything else aside when the job requires it. Finally, the third cause referred to different availability of aptitude at the high end: although women and men have similar average IQs, "there is a difference in the standard deviation, and variability of a male and a female population". In other words, men are over-represented at both the lowest and the highest levels, or according to the controversial quip of the time, men outnumber women among both idiots and geniuses. Summers may have made this last affirmation, which raised fiery reaction, based on the findings of several psycho-cognitive studies. In particular, Nowell and Hedges (1995) used data from six large scale surveys between 1960 and 1992 to investigate sex differences in intellectual abilities through mental test scores, and found that males outnumbered females substantially among high-scoring individuals (typically in mathematics and science). Afterwards, they confirmed that males are overrepresented in the upper tail regions of the composite math and science score distributions, using seven surveys representative of the Unites States twelfth grade student population and the National Assessment of Education and Progress (NAEP) long term trend data (Nowell and Hedges, 1998). Since top scientists are drawn disproportionately from people at the highest levels, Summers stated, then this is another possible factor in determining the diverse representation of women and men in high-end science. The authors of this study did not succeed in tracing research contributions that can scientifically support or refute this last thesis. The present study intends to furnish new evidence which could contribute to filling this gap.

In particular, with reference to the Italian academic system, this study proposes to verify if the higher average male performance in research, revealed by Abramo et al. in a preceding study (2007), can be largely ascribed to the subpopulation of the academic universe known as "star scientists": these are the scientists, to which Summers clearly referred, who stand out for the quantity and quality of their scientific production. The scope of the work is to compare and contrast the sex differences in research efficiency of star scientists with that of the rest of the academic population. In so doing, we will verify if there is: i) a higher concentration of men among star scientists, and ii) a higher performance of male star scientists with respect to female star scientists.

Unlike other investigations on this theme, the analysis will refer to an entire national academic population, that of all Italian universities, rather than to a sample of such a population, thus avoiding the potential limits of inferential analysis. The study, since it is conducted on the bases of detailed sectorial data, goes beyond the quantification of general phenomena to allow the highlighting of specific sectorial characteristics. In the remainder of this report, Section 2 defines the field of observation and the performance indicators to be applied. Section 3 gives a sketch of the characteristics seen in the field of observation and presents the results of the investigation concerning the issue at the source of its inspiration. Section 4 closes with conclusions and further considerations by the authors.

## 2. Data set and methodology

The proposed analysis is based on data extracted from the Italian Observatory of Public Research (ORP), which is maintained by the author's home research laboratory. The observatory is based on source data from the CD-rom version of the Thomson Reuters Science Citation Index (SCI™)[2]. Before elaborating data, we had to identify and unify the different ways in which the same organization was reported in the SCI™ "address" field for the articles[3]. Then, through a "disambiguation" algorithm, each publication[4] was attributed to its respective academic authors, with a margin of error of around 2%. This approach, although still a highly challenging technique, represents a remarkable novelty compared to other published research, as it permits overcoming the ranking distortions seen in the typical analysis conducted at higher levels of aggregation (Abramo and D'Angelo, 2007). It should also be noted that, by regulation, Italian university personnel are subdivided and assigned to specific scientific disciplinary sectors (SDS). Since the bibliometric measurements from the ORP concern single scientists, it was thus possible to carry out comparisons between those identified as belonging to a single SDS.

---

[2] The consideration of scientific journal publication as the sole output of research, excluding other recognized outputs such as proceedings, monographs, patents or prototypes, receives ample justification in the literature. A reaffirmation is that in the first national evaluation of research for Italy (VTR-CIVR, 2006), which took place during the same triennium under examination in this study, journal articles represented a minimum of 85% and a maximum of 99% of all products submitted for evaluation by universities; in seven areas of the eight technical and scientific areas considered, the incidence of publications was over 90% of the total products presented.
[3] For details see Abramo et al., 2008.
[4] Articles and reviews.

The field of observation is thus constituted of research personnel employed in Italian universities during the 2001 to 2003 triennium, belonging to the "hard" sciences. The Italian academic system is specifically subdivided into 14 disciplinary areas (DA) embedding 370 scientific disciplinary sectors. The analysis conducted here is concentrated on 8 areas[5], which in turn include 183 SDS. For analytical purposes, scientists that did not hold some position throughout the entire period were excluded from the observation. The study also excluded all scientists who, for whatever reason, changed their SDS during the triennium. This last step is in consideration of the difficulty of tracing exact individual identity, with resulting potential errors in the attribution of publications, due to frequent homonyms in the name and initials of authors. In total, these motives led to the exclusion of 3,780 individuals from the universe of 32,816 scientists. Finally, for those professionals who changed rank as a result of career advancement during the triennium, the analysis attributes the highest academic rank obtained in 2003, the final year of observation. The population thus used to conduct the analysis is illustrated in Table 1, as distributed by sex and academic rank. The average age of assistant professors is 44; of associate professors is 51; and of full professors is 57.

In the triennium under observation it results that 17,857 scientists published at least one publication[6], representing 61.5% of Italian academic research personnel[7]. The sex and rank allocation of the active academic scientists is illustrated in Table 2. The data do not differentiate in a significant manner between men and women: among women, 38.6% result as being inactive, compared to 38.5% of men. Still, when the data are disaggregated by academic rank and reconsidered, a meaningful gap appears, caused by the fact that women are primarily present in the rank of assistant professor, being the less "active" rank, as seen in Table 1. Among full professors, the proportion of active women is 1.1% higher with respect to that of men. For associate professors the difference rises to 1.5%, and for assistant professors as high as 5.2%.

[Table 1]
[Table 2]

Concerning the investigation of the above population, the scientific performance of each scientist was measured through application of the following indicators:
- Output (O): total of publications authored by the scientist in the survey period;
- Fractional Output (FO): total of the contributions made by the scientist to the publications, with "contribution" defined as the reciprocal of the number of co-authors of each article;
- Scientific Strength (SS): the weighted sum of the publications produced by the scientist, the weights for each publication being equal to the normalized impact factor[8] of the relevant journal[9];
- Fractional Scientific Strength (FSS): analogous to Fractional Output, but initiating from Scientific Strength.

---

[5] Mathematics and computer sciences, Physics, Chemistry, Earth sciences, Biological sciences, Medical sciences, Agriculture and veterinary sciences, Industrial and information engineering.
[6] Co-authored publications are attributed to each single author.
[7] Such low percentage of active academic scientists may be surprising to people not familiar with the Italian academic system, It can be explained by the lack of competition among universities, lack of research evaluation programs and lack of appropriate incentives. Also, the CD-rom version of SCI, our data source, indexes less journals, and thus less authors, than the Thomson-Reuters Web of Science.

Clearly, differing sets of star scientists are seen in correspondence with each indicator, each set being composed of individuals with performance among the top 10% within the scientific sector with which they are affiliated.

## 3. Results

The present work draws on conclusions of a preceding investigation by the same authors concerning the differences in research performance of men and women (Abramo et al., 2007). The study, referring to the data set noted above, revealed an average male productivity higher than that of women for all indicators of performance (although with important sectorial differences to be observed). In terms of Output, the average performance of males overall is higher than that of women (+16.5%), as it also is in each of the three single academic ranks analyzed (+13.3% for full professors, +12.3% for associate professors, +17.5% for researchers). From the same analysis, the performance gap seems to widen when the qualitative dimension is taken in consideration: in terms of Scientific Strength, differences are +19.7% for full professors, +15.9% for associate professors, +20.2% for researchers). It further emerged that:
- In each of the three ranks of the academic sector, the percentage of unproductive males is higher than that of women (32,1% vs 31% for full professors; 40,4% vs 38,9% for associate professors; 45,9% vs 40,7% for assistant professors). The overall average would seem to give a very slight indication to the contrary, but this is clearly due to the higher concentration of women in lesser academic ranks, within which a higher percentage of unproductive scientists are contained.
- More female than male scientists rank in the lowest levels of productivity. The contrary occurs for the highest levels of performance: this fact is again linked in part to the varying distribution of sex among academic ranks.
- The performance gap between the sexes seems to decline with career progress. This could in part be attributed to the effect of maternity leave[10], it being logical to expect that, for reasons of age, maternity and the attendant leave should occur more frequently among the lesser ranks in a university career.

Such results accord, in particular, with Lemoine (1992), who reveals that average productivity difference is also accompanied by diversity in distribution of productivity by sex: the concentration of women among those who publish a single article is higher than that of men, while it is lesser among star scientists. The particular characteristics of the community of "star scientists" have not been the object of significant investigation[11]. However it is possible that the primary contribution to the differential between the sexes can be attributed to exactly this part of the population, as the remainder of this article will attempt to verify.

---

[8] Normalization involved the transformation of the absolute value to percentile rank, based on the distribution of Impact Factor for all the journals in a given sector. In effect, the distribution of Impact Factor of journals is remarkably different sector by sector. The normalization makes it possible to narrow distortions inherent in measurements performed over different sectors.

[9] The Impact Factor represents a quantitative measure of the prestige of the journal. The authors are aware of the intrinsic limitations of such approximation, as well as of the recommendations contained in the literature on this issue (Moed and Van Leeuwen, 1996; Weingart, 2004). However, as their purpose was not that of providing an absolute ranking of the surveyed scientists, rather to compare rankings, the authors decided to present elaborations also on the basis of the Impact Factor measures.

[10] For reasons of privacy relative to maternity leave and all other leaves of absence, it was not possible to calculate research productivity in terms of the effective work time dedicated to the activity.

The population described in the above data set (Section 2) was divided in two complementary sub-groups. The first was composed of the so-called "star scientists". In each of the 183 SDS considered, the star scientists were identified as those located in the top 10% of the rankings of scientific performance. Obviously, according to the performance indicator selected, the precise top 10% identified as star scientists can be the subject of some variation. The second sub-group is composed of the remaining population, being all scientists with a performance less than the top 10%.

In the following section we will delineate the sex profile of star scientists, recognizing their salient differences with respect to the average character of the entire population (seen in the preceding study). Successively we will proceed to compare and contrast the sex differences in research performance between the two sub-groups.

*3.1 The "star scientist" profile*

In terms of Output, the subpopulation identified as star scientists consists of a total of 2,135 individuals. This number represents approximately 12% of the total population[12]. Their scientific production, averaged among the disciplinary areas, amounts to 35% of the total. A minimum of 23% is seen for SDS VET/07 (veterinary pharmacology and toxicology) and a maximum of 57% for SDS ING-IND/11 (environmental-technical physics). The distribution of star scientists by academic rank and sex is shown in Table 3: this subpopulation consists of 1,825 men (85.5% of total) and 310 women. Since the share of men in the overall academic population is 74.7%, this means that the concentration[13] of men among star scientists is 1.14, precisely twice that of the concentration of women among star scientists.

[Table 3]

In terms of academic rank, 57.8% of the total of star scientists is composed of full professors, with 28.1% being associate professors and 14.5% being assistant professors. The data thus suggest quite clearly that the star scientist is typically a male full professor. The index of concentration, equaling 1.60, indicates that the relative frequency of this profile among star scientists is over 60% greater than the frequency of the same profile in the entire population. It can be noted that no "dominant" academic rank emerges for female star scientists: 38% (117 of 310) assume an intermediate academic rank (associate professor), compared to 35% in a higher academic rank (full professor) and 28% in the lesser academic rank (assistant professor). For the other sex the situation is remarkably different: 61.8% of male star scientists are full professors, compared to 26.0% as associate and 12.2% as assistant professors. Analysis conducted by disciplinary areas (Table 4) confirms the indications that emerge from the general level, though with some meaningful differences. In particular, the indexes of concentration indicate that, among star scientists, women in the rank of full professor are more concentrated than men in three DA: agricultural and veterinary sciences, physics, and chemistry (however, in this last case the difference to the concentration of men is very slight). A further very interesting observation is that, in the DA of industrial and information engineering, the indexes of concentration do not provide any significant depiction of the "star scientist": different than in the other DA, there is no recognizable depiction, with certainty, of a particular combination of sex or academic rank.

---

[11] The term "star scientist" was coined by Zucker and Darby (1996) with reference to the specific scientific sector of biotechnology, with particular reference to the role of gatekeeper between the worlds of research and industry.

[12] In using percentiles to identify star scientists (in this case related to Output), we include all those on the threshold, leading to inclusion of a number greater than 10% of the total. Nine sectors were excluded from the investigation since they either presented a limited number of scientific researchers (less than five) or a case of a homogenous level of scientific production (which rends it impossible to individuate star scientists).

[13] The Concentration Index is a measure of association between two variables based on frequencies data and varying around the neutral value of 1. The value of 1.14 derives from the following ratio: (1,825/2,135)/(13,342/17,857), percentage of male star scientists divided by percentage of male scientists.

[Table 4]

Using the Fractional Scientific Strength (FSS) as the indicator for identification of top performers (Table 5), the results do not change much, although the maximum value of concentration index diminishes, and thus the significance of the results is also somewhat diminished. The most interesting variations concern mathematics and computer sciences and also industrial and information engineering. In mathematics and computer sciences it emerges that the relative frequency among male top scientists for assistant professors (1.20) is very close to that for full professors (1.25). In the engineering DA, the maximum concentration of star scientists occurs in the female associate professor rank (1.48), which significantly outdistances the value for males in the same academic rank (1.16).

[Table 5]

The substantially similar nature of the data just seen in Tables 4 and 5 leads to inductive reasoning that the phenomena examined in this study remain substantially invariant, regardless of the type of performance being considered. However, as we will better see in the following sections, it seems that it is the quantitative dimension that better discriminates the difference between the two populations of star scientists under comparison.

*3.2 Sex differences in research productivity of star scientists*

Analysis of the average frequency of articles published in a year indicates that, at a general level, female star scientists are primarily concentrated in the lesser levels of productivity (Figure 1). Specifically, 28.1% of women top scientists produce less than three publications per year, while 23.2% of the males register this lower level of productivity. Meanwhile, 8% of males register an average scientific production superior to 10 articles per year, while only 2% of females reach this level. From lowest to highest frequency of production (viewing from left to right), there is an evident reversal of the sexes, as indicated by the bars of the histogram in Figure 1. Further, the index of skewness, again for distribution of Output, which results as 0.726 for women and 2.206 for men, suggests a substantially normal distribution for males but one that is highly asymmetrical and heterogeneous for female star scientists.

[Figure 1]

Such results must be interpreted with the necessary caution, since research productivity is itself clearly non-homogenous by scientific sector, just as the distribution of men and women is also non-homogenous. To accommodate for this non-homogenous situation, the comparison in performance between the sexes was further conducted at the level of single SDS. In addition, since production by academic rank is also non-homogeneous (Abramo et al. 2007) as is the division of the sexes by rank (though the distribution of star scientists is concentrated, sector by sector, at the higher levels of performance), this guided the authors to again conduct analysis differentiated by academic rank.

To identify performance differences between females and males (F, M), the authors used a calculation of average general performance ($\overline{P}g_k$) for star scientists of sex $g$ and academic rank $k$ as follows:

$$\overline{P}g_k = \frac{1}{Sg_k} \sum_{j=1}^{n_{SDS}} \frac{\overline{P}g_{jk}}{\overline{P}_{jk}} \cdot Sg_{jk} \qquad [1]$$

with:

$\overline{P}g_{jk}$ = average performance of star scientists of sex $g$ (F,M) and role $k$, in sector $j$

$\overline{P}_{jk}$ = average performance of star scientists of role $k$ in sector $j$

$Sg_{jk}$ = number of star scientists of sex $g$ and role $k$, in sector $j$

$Sg_k$ = total of star scientists of sex $g$ and role $k$

$n_{SDS}$ = number of sectors under observation

Table 6 presents the absolute values and percentage differences of $\overline{P}g_k$ for the two sexes. In terms of Output, the higher average performance of male star scientists is quite evident, being +10.3% among full professors, +7.1% among associate professors and +3.9% among assistant professors. The situation remains constant when examining the qualitative dimension of performance: in terms of Scientific Strength (SS) for full professor star scientists, males show a performance 9.6% greater than that of females, while among both associate and assistant professors the difference is 3.3%. Fractional Scientific Strength (FSS), which is the indicator that gathers all dimensions of evaluation (quantitative, qualitative and contributive), shows further heightening of the difference in favor of men, in all three academic ranks: the averages are +13.4% higher for full professors, +13.5% for associate and 7.5% for assistant professors. It is evident from this inter-rank analysis that the gap between men and women tends to increase with career progress. The right hand column of Table 6 indicates that, proceeding to an aggregation of the data, normalized by academic rank and weighted for presence relative to the academic ranks, male performance is invariably superior to that by females.

[Table 6]

Comparisons between average performance of males and females could potentially be influenced by peaks in the performance registered by single scientists. In order to better compare the placement of single individuals in the performance rankings, this study applies the "causal variables sequence criterion". This criterion is based on calculation of the "distance" from the boundary condition of maximum difference in performance between the sexes[14] based on the distribution, within each sector, of rankings for each of the 2,163 individuals in the data set[15].

Results for each single sector were then successively aggregated by DA and gathered in Tables 7, 8 and 9. The tables show, for each academic rank, the sex that registers the "highest" position (that which approaches the ideal situation most closely) both overall and within each disciplinary area, indicator by indicator[16].

For full professors, the causal variables sequence criterion clearly indicates that average performance of male star scientists for Output, Fractional Output and Fractional Scientific Strength is not less than (is greater than or equal to) that of women, in all disciplinary areas (Table 7). However, the analysis of the data set for the fourth indicator, Scientific Strength, shows that in three disciplinary areas out of eight, the situation is the inverse. The three areas concerned are agricultural and veterinary sciences, earth sciences and biological sciences.

[Table 7]

For the rank of associate professor (Table 8) the situation shown by analysis is essentially the same as that for full professors: the overall contrast between the sexes remains in favor of males, for all the indicators. However, when distinguishing the data by disciplinary area, women demonstrate performance not less than that of men, for both Output and Scientific Strength, in the DA of agricultural and veterinary sciences. In the DA of chemistry, the average female ranking is not less than that of men for Fractional Output and Fractional Scientific Strength. Finally, both earth sciences and physics show a situation where one of the four indicators registers a performance of female star scientists not less than that of their male colleagues.

Lastly, focused on assistant professors, it emerges that in this rank women star scientists quite frequently achieve higher results than men (Table 9). In terms of Output, the average performance of women is not less than that of men in five disciplinary areas out of eight, and also at the overall level. For Fractional Output, however, the male sex always prevails. This indicates that, more than men, women tend to publish in co-authorship.

[Table 8]
[Table 9]

In general, the "inter-rank" analysis highlights that the gap between the sexes tends to increase with career progression, as again seen in the preceding subsection. In fact, considering the 32 situations represented by the combination of the eight DA and four indicators considered, female performance is not less than that of males in 13 cases for assistant professors, lessening to six cases for associate professors and three cases for full professors.

---

[14] "Maximum difference" is intended to mean the situation in which the highest performing woman (or man) is located lower than the lowest performing man (or woman).
[15] For the details concerning this criteria see Abramo et al., 2007.
[16] Note that the indicator considered in each case is also that utilized to identify the relevant group of star scientists.

*3.3 Sectorial analysis*

The comparison between star scientists of the two sexes was also continued at the more detailed level of the individual SDS that falls within each disciplinary area. For each DA, Tables 10, 11 and 12 indicate the number of sectors in which the average rank of female star scientist is not less than that of males. Such comparison is obviously possible only in sectors where there is representation of star scientists from both sexes, and in consequence the total number of sectors under comparison in the tables varies with academic rank and with each indicator[17].

Considering the population of "full professor star scientists" (Table 10), the average percentile rank for Output by women is not less than that of men in 24 sectors out of 52 (46.2%). Results that are not dissimilar are also obtained when considering other performance indicators: for example in Fractional Scientific Strength, in the 43 sectors in which a comparison is possible, women prevail in 11 cases (25.6%). Examining the data, there are also some evident differences among the disciplinary areas: in agricultural and veterinary sciences the comparison frequently sees women prevail (in 5 sectors out of 9 for Output, in 4 out of 7 for Scientific Strength). The same occurs in biological sciences: in terms of Output, the average performance of female star scientists is not less than that of males in 9 sectors out of 13, also true for women in 6 out of 11 sectors for Scientific Strength. On the contrary, in medical sciences (the DA which registers 31% of the entire field of observation, in terms of number of scientists) the number of sectors in which women prevail is truly narrow, independently of the indicator. For the other disciplinary areas the comparison between the sexes is limited to few sectors and it is difficult to draw general conclusions.

[Table 10]

Table 11 presents data from the comparison of star scientists in the associate professor rank. The average percentile rank for Output by women is not less than that of men in 24 sectors of 64 (37.5%); in terms of Scientific Strength this occurs in 29 sectors of 61 (47.5%).

[Table 11]

With respect to single DA, in industrial and information engineering a comparison in favor of women emerges more frequently, concerning up to 7 to 9 sectors with the application of the various indicators: notably, for Output and Scientific Strength a performance of women greater than that of men is seen in three sectors, and the same is seen in two sectors for both Fractional Output and Fractional Scientific Strength. In agricultural and veterinary sciences, the prevailing tendency of women seen in the analysis of full professors is again confirmed: the contrast of average percentile rank sees women prevail in the majority of cases, in particular in eight sectors out of 11 for Output and in six of 9 sectors for scientific force. Biological sciences registers a case of virtual parity, while in medical sciences the occurrence of higher performance by women is still limited, but more noticeable than among full professors, reaching a level

---

[17] In effect, as the indicator used to identify top performers changes, the set of star scientists subjected to analysis also changes, and as a consequence the number of sectors in which the performance between men and women can be confronted also varies.

of three sectors out of 13 for Output and five of 13 for Scientific Strength.

The analysis conducted for assistant professors (Table 12) does not reveal significant differences when compared to the other academic ranks. In terms of Output, for example, the number of sectors in which female star scientists prevail arrives at 22 out of 41 among assistant professors (53.7%) compared to the 24 of 52 sectors for full professors (46.2%). The situation does not change concerning Scientific Strength – women register an average performance superior to men in 15 sectors of 34 among assistant professor star scientists, compared to 15 of 42 among full professor star scientists.

[Table 12]

### *3.4 Sex differences among star scientists as compared to the rest of the population*

We can now verify if the higher average performance by male researchers can be ascribed in a preponderant manner to the effect of "star scientists". The performance gap between the sexes in the subpopulation of star scientists will be contrasted to that of complementary subpopulations and of the entire population. Since the range in performance of the two subpopulations is very different, steps were first taken to normalize the individual performance with respect to the pertinent population: the disciplinary sector of affiliation for the scientist. In this regard, the following formula was applied:

$$In_{igs} = \frac{I_{igs} - I_{MIN_S}}{I_{MAX_S} - I_{MIN_S}} \quad [2]$$

with:

$In_{igs}$ = normalized performance of scientist *i* of sex *g* and sector *s*

$I_{igs}$ = absolute performance of scientist *i* of sex *g* and sector *s*

$I_{MIN_S}$ = minimum observed value of performance of scientists in sector *s*

$I_{MAX_S}$ = maximum observed value of performance of scientists in sector *s*

In this manner, individual levels of performance always fall within two extreme values (0 and 1), and can be readily subjected to comparisons between diverse populations. The percentage gaps in favor of males that result from the calculation of normalized performance are seen in Table 13.

[Table 13]

The results show unequivocally that sex differences in performance of the entire population of Italian academic population are determined, in a preponderant manner, by star scientists.

For full professors, while we see an average production (Output) of males that is superior to that of females by 19.7% among the total university population, we are confronted with a difference of a full 44.3% among star scientists, and at the same moment a gap of only 2.2% for the rest of the population. The result reached is substantially invariant for all four indicators considered. Indeed, in terms of Fractional Scientific Strength for male and female star scientists, the performance gap in favor of males results as 35.2%, while in the rest of the population the direction of the difference is actually inverted.

Analysis for the associate professor rank offers no significant differences. For assistant professors, once again, the sex differences seem highly relevant among star scientists (actually reaching +117.5% in terms of Fractional Scientific Strength) and less in the rest of the population, though still remaining significant (+9.8% for FSS).

The inter-rank analysis conducted for subpopulation complementary to the star scientists seems to indicate that here, sex differences tend to reduce with career progress: in terms of Output, the performance gap equals 7.6% among assistant professors, 4% among associate professors and 2.2% among full professors. This result corresponds with that of the preceding analysis conducted for the entire population of authors (Abramo et al., 2007) and can very likely be explained by the effect of maternity leave: such leave, for reasons of age, is significantly more frequent in the lesser university career ranks. The same analysis for the star scientists shows a reverse tendency. Presumably the intensity of maternity among star scientists is lesser than that of the complement, although privacy regulations do not permit us the data to demonstrate the point conclusively.

The same type of analysis was conducted with reference to the average percentile rank, normalized with respect to the amplitude of distribution of the sectorial values. The resulting differences in measurements are indicated in Table 14.

[Table 14]

It can be noted that validity is retained for all the observations made up to this point: for full professors, the sex difference in percentile ranking of performance (in favor of males) is 6% for Output and Scientific Strength and 9.7% for Fractional Scientific Strength. For the complementary population, however, the sex difference is inverted: the difference in performance is in favor of women, though the absolute value is quite close to zero. This tendency is also true for associate professors: the gap in favor of males is evident for the population of star scientists, while tending to reduce towards zero for the rest of the population. For the assistant professors, in contrast, at least in terms of Output, the gap between male and female star scientists disappears. The inter-rank analysis conducted for the cohort of scientists not included in the "top" category also confirms the previously-seen trend of a reduction in sex gap with career progress: in terms of Output, the performance difference equals 3.8% among assistant professors, lessens to 1.5% among associate professors and is negative for full professors (the same trend is also seen in the results for all the other indicators used in the analysis). As noted above, star scientists demonstrate a different tendency, with the sex performance gap actually increasing with career progression for some indicators (Output and Scientific Strength).

## 4. Conclusions

Literature dedicated to analyzing performance differences between the women and the men employed in research seems to agree that, factually, males publish more than females. However regarding the causes of such differences, the literature provides very few contributions, which at times cannot be completely reconciled. The current authors are among those who, through a previous study of the entire population of Italian academic scientists (Abramo et al. 2007), reveal an average productivity by males which is higher than that of women, although with the presence of some important sectorial differentiations.

The present work was intended to verify if such performance differentials could be ascribed in a preponderant manner to the subpopulation of university academics that is remarked for its research efficiency: the so-called "star scientists". The series of analyses conducted confirm this hypothesis. Above all:

- males have a concentration among star scientists that is double that of women;
- female star scientists are concentrated in the lower levels of productivity with respect to their male colleagues;
- the scientists that produce over 10 articles per year include 8% of the men, but only 2% of the women;
- in general, the performance of male star scientists both in terms of average and ranking of single individuals is superior both at the general level and in each of the three academic ranks considered separately.

The sex difference in performance among star scientists contributes in a preponderant manner to determining the sex gap for the entire population of all academic scientists. For the remaining 90% of the population the sex differences are practically inexistent and even reverse to be in favor of women, in the case of full professors. Evidently, to obtain levels of scientific production such as those of a star scientist, the time and energy required for research activities are notably superior to the average, and imply an overwhelming dedication to work. This may constitute an element of discrimination against those individual women who have a greater interest or commitment, with respect to men, to balancing professional and family life, with the difficulty of the respective time requirements. Another possible explanation of the results is suggested by the area of the literature that indicates the existence of a psycho-cognitive gap between the sexes, especially in the technical-scientific disciplines. However, it is not the present intention of the authors to support one hypothesis in particular, neither to offer particular investigation of the determinants of the phenomena observed here.

Another important result of the study obtains from the "inter-rank" analysis. Among the star scientists, the difference in performance in favor of men tends to augment with career progression; in the rest of the population it tends to diminish, even up to the point, at the level of full professor, of inverting to the favor of women. This contrasting evidence could again be explained (only in logical-deductive terms, given the regulatory considerations that block access to data) by a different impact of the motherhood role on these two populations. The distribution of age relative to Italian academic ranks indicates that it would above all be female assistant professors that are impacted by the interruption of scientific work for child-bearing. This could in turn bear upon on the average performance of the category and on the contrast with the male complement. On the other hand, as previously stated, it is also probable that the work experience for star scientists, independently of rank, is difficult to reconcile with the fact of maternity, and could indicate a substantial "equality of conditions" among male and female star scientists, in particular at the lower career levels.

The exhaustive nature of the field of observation, with the capacity to base the investigation on the overall data rather than on samples, constitute new elements with respect to the literature on the subject, and rend the results more robust. Though the present analysis concerns only the academic reality in Italy, this is certainly a national research system that is relevant in size and significance. The question remains open concerning the causes of the facts placed in evidence. The intent of the authors has been to furnish further evidence to their colleagues in psycho-cognitive studies and sociology of science, so that from their respective bases they may definitively resolve the "productivity puzzle".

**References**

Abramo, G. and D'Angelo, C.A. (2007), Measuring Science: Irresistible Temptations, Easy Shortcuts and Dangerous Consequences, *Current Science* 93(6): 762-766.

Abramo, G., D'Angelo C.A. and Pugini, F. (2008), The Measurement of Italian Universities' Research Productivity by a non Parametric-Bibliometric Methodology, *Scientometrics* 76(2).

Abramo, G., D'Angelo, C.A. and Caprasecca, A. (2007), Gender differences in research productivity: a bibliometric survey on the Italian academic system, forthcoming in *Scientometrics*

Cole, J. R. and Zuckerman, H. (1984), The productivity puzzle: persistence and change in patterns in publication of men and women scientists, *Advances in motivation and achievement* 2: 217-258.

Etzkowitz, E., Kemelgor, C. and Uzzi, B. 2002. *Athena Unbound: The Advancement of Women in Science and Technology*, Cambridge University press.

Fox, M.F. (1983), Pubblication productivity among scientists: a critical review, *Social Studies of Science* 13(2): 285-305.

Fox, M.F. (2005), Gender, family characteristics, and publication productivity among scientists, *Social Studies of Science* 35(1): 131–150.

Fox, M.F. and Mohapatra S. 2007. Social-Organizational Characteristics of Work and Publication Productivity Among Academic Scientists in Doctoral-Granting Departments. *Journal of Higher Education*, 78(5): 543-571.

Hyde, J.S. and Linn, M.C. (1988), Gender differences in verbal ability: A meta-analysis, *Psychological Bulletin* 104: 53–69.

Hyde, J.S., Fennema, E. and Lamon, S.J. (1990), Gender differences in mathematics performance: A meta-analysis, *Psychological Bulletin* 107: 139–155.

Leahey, E. (2006), Gender differences in productivity: research specialization as a missing link, *Gender and Society* 20(6): 754-780.

Lee, S. and Bozeman, B. (2005), The Impact of Research Collaboration on Scientific Productivity, *Social Studies of Science* 35(5): 673-702.

Lemoine, W. (1992), Productivity patterns of men and women scientists in Venezuela, *Scientometrics* 24(2): 281-295.

Linn, M.C. and Petersen, A.C. (1985), Emergence and characterisation of sex differences in spatial ability: A meta-analysis, *Child Development* 56: 1479–1498.

Long, J.S. (1987), Problems and Prospects for Research on Sex Differences in the Scientific Career, In L.S. Dix (ed.) *Women: Their Underrepresentation and Career Differentials in Science and Engineering* (National Academy Press).

Long, J.S. (1992), Measure of Sex Differences in Scientific Productivity, *Social Forces* 71(1): 159-178.

Mauleón, E. and Bordons, M. 2006. Productivity, impact and publication habits by gender in the area of Materials Science, *Scientometrics*, 66 (1): 199 – 218.

Moed, H.F. (2002), The impact factors debate: the ISI's uses and limits, *Nature* 415: 731-732.

Nowell, A. and Hedges, L.V. (1995), Sex differences in mental test scores, variability and numbers of high-scoring individuals, *Science* 269: 41-45.

Nowell, A. and Hedges, L.V. (1998), Trends in Gender Differences in Academic Achievement from 1960 to 1994: An Analysis of Differences in Mean, Variance, and Extreme Scores, *Sex Roles* 39(1/2).

Palomba, R. and Menniti, A. (Eds) 2001. *Minerva's Daughters*. Institute for Research

on Population and Social Policies. Rome, Italy.

Pripiċ, K. (2002), Gender and productivity differentials in science, *Scientometrics* 55(1): 27-58.

Stack, S. (2004), Gender, Children and research Productivity, *Research in Higher Education* 45(8).

Summers, L.H. (2005), Remarks at NBER Conference on Diversifying the Science & Engineering Workforce, (Cambridge, Mass. - January 14), http://www.president.harvard.edu/speeches/2005/nber.html.

Voyer, D., Voyer, S. and Bryden, M.P. (1995), Magnitude of sex differences in spatial ability: A meta-analysis and consideration of critical variables, *Psychological Bulletin* 117: 250–270.

VTR-CIVR (2006), *VTR 2001-2003. Risultati delle valutazioni dei Panel di Area,* http://vtr2006.cineca.it/

Weingart, P. (2004), Impact of bibliometrics upon the science system: inadvertent consequences?, In H. F. Moed, W. Glänzel, U. Schmoch (eds.), *Handbook on Quantitative Science and Technology Research*, (Dordrecht, The Netherlands, Kluwer Academic Publishers).

Xie, Y. and Shauman, K.A. (1998), Sex differences in research productivity: new evidence about an old puzzle, *American Sociological Review* 63: 847-870.

Xie, Y. and Shauman, K.A. (2003) *Women in Science: career processes and outcomes* (Harvard University Press, Cambridge, Massachusetts, and London, England).

Zainab, A.N. (1999), Personal, academic and departmental correlates of research productivity: a review of literature, *Malaysian Journal of Library & Information Science* 4(2): 73-110.

Zucker, L.G. and Darby, M.R. (1996), Star scientists and institutional transformation: Patterns of invention and innovation in the formation of the biotechnology industry, *Proceedings of the National Academy of Science of USA* 93: 12709–12716.

|        | Full professors | Associate professors | Assistant professors | Total |
|--------|----------------|---------------------|---------------------|-------|
| Male   | 8686 (88.7%)   | 7596 (73.4%)        | 5401 (60.7%)        | 21683 (74.7%) |
| Female | 1102 (11.3%)   | 2758 (26.6%)        | 3493 (39.3%)        | 7353 (25.3%) |
| Total  | 9788 (33.7%)   | 10354 (35.76%)      | 8894 (30.6%)        | 29036 |

*Table 1: Distribution of the population of Italian university research staff by sex and academic rank (parentheses indicate the percentage active in publication); data set 2001 to 2003*

|        | Full professors | Associate professors | Assistant professors | Total |
|--------|----------------|---------------------|---------------------|-------|
| Male   | 5895 (67.9%)   | 4526 (59.6%)        | 2921 (54.1%)        | 13342 (61.5%) |
| Female | 760 (69.0%)    | 1684 (61.1%)        | 2071 (59.3%)        | 4515 (61.4%) |
| Total  | 6655 (68.0%)   | 6210 (60.0%)        | 4992 (56.1%)        | 17857 (61.5%) |

*Table 2: Distribution of "active" Italian university research staff by sex and academic rank; data set 2001 to 2003*

|                      |               | Male  | Female | Total |
|----------------------|---------------|-------|--------|-------|
| Full professors      | Number        | 1,127 | 107    | 1,234 |
|                      | Incidence     | 91.3% | 8.7%   | 57.8% |
|                      | Concentration | 1.60  | 1.18   |       |
| Associate professors | Number        | 475   | 117    | 592   |
|                      | Incidence     | 80.2% | 19.8%  | 28.1% |
|                      | Concentration | 0.58  | 0.88   |       |
| Assistant professors | Number        | 223   | 86     | 309   |
|                      | Incidence     | 72.2% | 27.8%  | 14.5% |
|                      | Concentration | 0.35  | 0.64   |       |
| Total                | Number        | 1,825 | 310    | 2,135 |
|                      | Incidence     | 85.5% | 14.5%  |       |
|                      | Concentration | 1.14  | 0.57   |       |

*Table 3: Sex and academic rank distribution for "star scientists" (as identified from Output); average data 2001 to 2003 for all "hard" sciences disciplinary areas, excluding civil engineering and architecture*

|                                       | Full professors || Associate professors || Assistant professors ||
|                                       | M    | F    | M    | F    | M    | F    |
|---------------------------------------|------|------|------|------|------|------|
| Industrial and information engineering | 1.25 | 0.96 | 1.03 | 1.00 | 0.84 | 0.34 |
| Agriculture and veterinary sciences    | 1.48 | 1.70 | 1.28 | 0.95 | 1.00 | 0.75 |
| Biological sciences                    | 1.80 | 1.06 | 0.87 | 0.54 | 0.57 | 0.32 |
| Chemistry                              | 1.55 | 1.58 | 0.71 | 0.44 | 0.66 | 0.32 |
| Earth sciences                         | 1.57 | 0.76 | 1.03 | 0.69 | 0.73 | 0.61 |
| Physics                                | 1.33 | 1.52 | 0.83 | 0.97 | 0.62 | 0.23 |
| Mathematics and computer sciences      | 1.54 | 0.91 | 1.02 | 0.67 | 1.01 | 0.50 |
| Medical sciences                       | 1.83 | 1.07 | 0.78 | 0.41 | 0.47 | 0.23 |

*Table 4: Indexes of concentration for star scientists (as identified from Output) by disciplinary area; period 2001 to 2003*

|                                       | Full professors || Associate professors || Assistant professors ||
|                                       | M    | F    | M    | F    | M    | F    |
|---------------------------------------|------|------|------|------|------|------|
| Industrial and information engineering | 1.11 | 0.82 | 1.16 | 1.48 | 0.64 | 0.51 |
| Agriculture and veterinary sciences    | 1.15 | 1.19 | 1.26 | 0.78 | 0.74 | 0.86 |
| Biological sciences                    | 1.94 | 0.87 | 0.90 | 0.48 | 0.57 | 0.33 |
| Chemistry                              | 1.53 | 1.35 | 0.76 | 0.50 | 0.92 | 0.42 |
| Earth sciences                         | 1.54 | 0.43 | 0.82 | 0.94 | 0.72 | 0.87 |
| Physics                                | 1.19 | 1.16 | 0.99 | 0.61 | 0.89 | 0.44 |
| Mathematics and computer sciences      | 1.25 | 0.81 | 1.07 | 0.51 | 1.20 | 0.40 |
| Medical sciences                       | 1.88 | 0.99 | 0.83 | 0.35 | 0.54 | 0.22 |

*Table 5: Indexes of concentration for star scientists (as identified by Fractional Scientific Strength) by disciplinary area; 2001 to 2003 period*

| Index | Sex | Full professors | Associate professors | Assistant professors | Total |
|---|---|---|---|---|---|
| O | M | 1.008 (+10.3%) | 1.013 (+7.1%) | 1.010 (+3.9%) | 1.010 (+7.2%) |
|   | F | 0.914 | 0.946 | 0.973 | 0.942 |
| SS | M | 1.007 (+9.6%) | 1.006 (+3.3%) | 1.010 (+3.3%) | 1.007 (+5.1%) |
|   | F | 0.919 | 0.974 | 0.978 | 0.958 |
| FO | M | 1.010 (+13.7%) | 1.009 (+5.0%) | 1.020 (+8.7%) | 1.011 (+8.7%) |
|   | F | 0.888 | 0.961 | 0.939 | 0.93 |
| FSS | M | 1.009 (+13.4%) | 1.022 (+13.5%) | 1.020 (+7.5%) | 1.014 (+11.1%) |
|   | F | 0.889 | 0.901 | 0.949 | 0.913 |

*Table 6: Average normalized performance ($\overline{P}g_k$) of star scientists subdivided by sex and academic rank*

| Disciplinary Area | O | SS | FO | FSS |
|---|---|---|---|---|
| Industrial and information engineering | M | M | M | M |
| Agriculture and veterinary sciences | M | F | M | M |
| Biological sciences | M | F | M | M |
| Chemistry | M | M | M | M |
| Earth sciences | M | F | M | M |
| Physics | M | M | M | M |
| Mathematics and computer sciences | M | M | M | M |
| Medical sciences | M | M | M | M |
| Total | M | M | M | M |

*Table 7: Disciplinary areas, indicating the sex with the best average performance, as identified by causal variables sequence criterion (data referring to full professor star scientists)*

| Disciplinary Area | O | SS | FO | FSS |
|---|---|---|---|---|
| Industrial and information engineering | M | M | M | M |
| Agriculture and veterinary sciences | F | F | M | M |
| Biological sciences | M | M | M | M |
| Chemistry | M | M | F | F |
| Earth sciences | F | M | M | M |
| Physics | M | M | F | M |
| Mathematics and computer sciences | M | M | M | M |
| Medical sciences | M | M | M | M |
| Total | M | M | M | M |

*Table 8: Disciplinary areas, indicating the sex with the higher average performance, as identified by causal variables sequence criterion (data referring to associate professor star scientists)*

| Disciplinary Area | O | SS | FO | FSS |
|---|---|---|---|---|
| Industrial and information engineering | F | M | F | F |
| Agriculture and veterinary sciences | F | F | M | M |
| Biological sciences | M | M | M | M |
| Chemistry | M | M | M | M |
| Earth sciences | F | M | F | M |
| Physics | F | F | M | M |
| Mathematics and computer sciences | F | F | F | M |
| Medical sciences | M | M | M | F |
| Total | F | M | M | M |

*Table 9: Disciplinary areas, indicating the sex with the higher average performance, as identified by causal variables sequence criterion (data referring to assistant professor star scientists)*

| Disciplinary Area | O | SS | FO | FSS |
|---|---|---|---|---|
| Industrial and information engineering | 1 of 2 | 0 of 2 | 0 of 1 | 0 of 1 |
| Agriculture and veterinary sciences | 5 of 9 | 4 of 7 | 3 of 8 | 3 of 7 |
| Biological sciences | 9 of 13 | 6 of 11 | 3 of 12 | 3 of 11 |
| Chemistry | 2 of 6 | 2 of 7 | 1 of 7 | 1 of 7 |
| Earth sciences | 0 of 2 | 1 of 1 | 0 of 1 | 0 of 1 |
| Physics | 1 of 4 | 0 of 3 | 0 of 4 | 1 of 4 |
| Mathematics and computer sciences | 3 of 5 | 1 of 3 | 2 of 3 | 2 of 4 |
| Medical sciences | 3 of 11 | 1 of 8 | 1 of 11 | 1 of 8 |
| Total | 24 of 52 | 15 of 42 | 10 of 47 | 11 of 43 |

*Table 10: Number of scientific disciplinary sectors in which the average percentile rank for female full professor star scientists is not less than that of males*

| Disciplinary Area | O | SS | FO | FSS |
|---|---|---|---|---|
| Industrial and information engineering | 3 of 9 | 3 of 9 | 2 of 7 | 2 of 9 |
| Agriculture and veterinary sciences | 8 of 11 | 6 of 9 | 2 of 5 | 3 of 7 |
| Biological sciences | 5 of 13 | 4 of 8 | 6 of 12 | 2 of 11 |
| Chemistry | 1 of 7 | 6 of 9 | 3 of 6 | 3 of 5 |
| Earth sciences | 2 of 3 | 2 of 3 | 1 of 2 | 0 of 1 |
| Physics | 2 of 3 | 1 of 3 | 1 of 3 | 0 of 2 |
| Mathematics and computer sciences | 2 of 5 | 2 of 7 | 3 of 5 | 4 of 6 |
| Medical sciences | 3 of 13 | 5 of 13 | 3 of 9 | 3 of 9 |
| Total | 26 of 64 | 29 of 61 | 21 of 49 | 17 of 50 |

*Table 11: Number of scientific disciplinary sectors in which the average percentile rank for female associate professor star scientists is not less than that of males*

| Disciplinary Area | O | SS | FO | FSS |
|---|---|---|---|---|
| Industrial and information engineering | 2 of 2 | 2 of 3 | 1 of 1 | 2 of 3 |
| Agriculture and veterinary sciences | 7 of 9 | 2 of 2 | 1 of 4 | 1 of 2 |
| Biological sciences | 3 of 7 | 2 of 6 | 3 of 8 | 0 of 7 |
| Chemistry | 1 of 5 | 3 of 4 | 1 of 2 | 1 of 6 |
| Earth sciences | 1 of 1 | 1 of 1 | 0 of 1 | 1 of 1 |
| Physics | 1 of 2 | 2 of 2 | 0 of 3 | 0 of 3 |
| Mathematics and computer sciences | 5 of 6 | 2 of 5 | 1 of 5 | 2 of 6 |
| Medical sciences | 2 of 9 | 1 of 11 | 2 of 6 | 3 of 8 |
| Total | 22 of 41 | 15 of 34 | 8 of 30 | 10 of 36 |

*Table 12: Number of scientific disciplinary sectors in which the average percentile rank for female assistant professor star scientists is not less than that of males*

| | Full professors | | | Associate professors | | | Assistant professors | | |
|---|---|---|---|---|---|---|---|---|---|
| Index | StS | Others | Total | StS | Others | Total | StS | Others | Total |
| O | 44.3% | 2.2% | 19.7% | 85.7% | 4.0% | 19.3% | 16.4% | 7.6% | 21.5% |
| SS | 25.1% | -2.9% | 22.5% | 39.3% | 0.8% | 22.6% | 29.9% | 7.7% | 19.7% |
| FO | 71.5% | -3.9% | 21.2% | 55.4% | 3.4% | 22.8% | 80.0% | 9.0% | 28.6% |
| FSS | 35.2% | -2.6% | 27.6% | 55.3% | 4.6% | 32.3% | 117.5% | 9.8% | 33.5% |

*Table 13: Percentage difference in performance, in favor of males, for the subpopulations of star scientist (StS), the rest of the population (Others), total population (Total), and for ranks as academic professional.*

| | Full professors | | | Associate professors | | | Assistant professors | | |
|---|---|---|---|---|---|---|---|---|---|
| Index | StS | Others | Total | StS | Others | Total | StS | Others | Total |
| O | 6.0 | -0.8 | 1.5 | 11.3 | 1.5 | 3.0 | -0.2 | 3.8 | 4.8 |
| SS | 6.1 | -1.6 | 1.6 | 7.3 | 0.5 | 2.3 | 4.2 | 3.2 | 4.1 |
| FO | 12.4 | -0.8 | 1.8 | 9.3 | 1.9 | 3.6 | 11.8 | 4.0 | 5.4 |
| FSS | 9.7 | -1.2 | 2.0 | 11.2 | 1.4 | 3.3 | 11.7 | 3.6 | 4.8 |

*Table 14: Difference in average percentile rank in favor of males for the subpopulations of star scientists (StS), the rest of the population (others), total population (total), and for rank as academic*

*professional; values indicated are percentages*

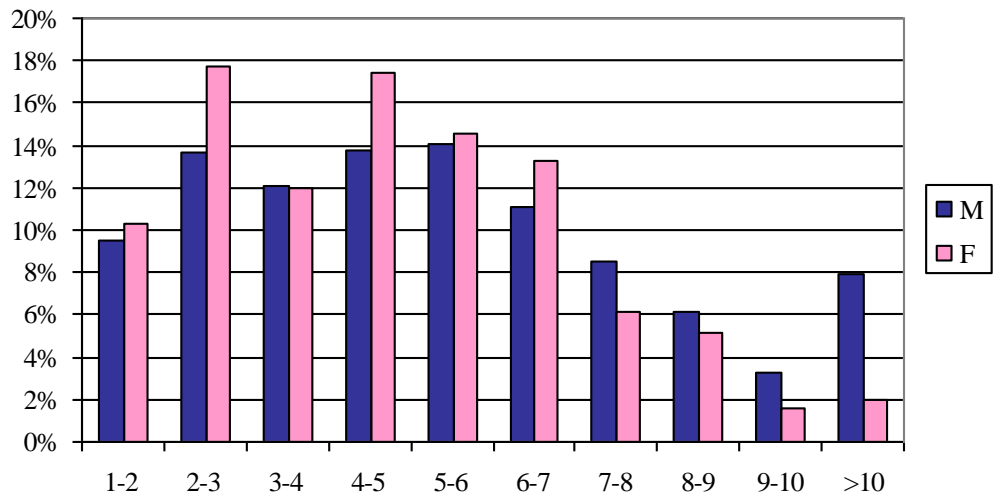

*Figure 1: Analysis of frequency of average annual scientific production (number of articles) by Italian academic star scientists; 2001 to 2003 period*